# Study of locally frozen magnetic field in a high-$T_c$ superconducting ceramic

S. I. Bondarenko, A. A. Shablo, and V. P. Koverya

*B. Verkin Institute for Low Temperature Physics and Engineering, National Academy of Sciences of Ukraine, pr. Lenina 47, Kharkov 61103, Ukraine*

The properties of a locally frozen (in a region of diameter 0.5 mm) magnetic field in a $YBa_2Cu_3O_{7-x}$ slab 0.5 μm thick are investigated as a function of the value of the excitation field, the regime of freezing, and the transport current through the sample. The first regime is cooling of the ceramic to 77 K in the excitation field with a subsequent turning off of the excitation field, and the second regime is cooling in the Earth's magnetic field with a subsequent turning on and off of the excitation field. At an excitation field up to 2000 A/m in these regimes two different types of macroscopic current vortex structures, which generate the frozen field, are formed. The local critical field of excitation when the vortex structure is formed in the second regime exceeds the uniform perpendicular critical field of the slab by a factor of 10 and equals 1700 A/m. On the other hand, the vortex structure of the first type can be formed by practically any weak excitation field, including fields smaller than the critical field of the vortex structure of the second type. In a representation of the ceramic as a Josephson medium, physical models of the two types of vortex structures are proposed which correspond most fully to the results of experiments. The displacement of the vortex structure of the first type upon the passage of transport current through the slab, as a result of the action of the Lorentz force on that structure, is registered. This makes it possible to calculate the pinning force $F_p$ and to estimate the value of the viscosity $\eta$ for the motion of such a vortex structure in the ceramic: $F_p = 6 \times 10^{-8}$ N, $\eta = 6 \times 10^{-5}$ kg/s.

## I. INTRODUCTION

Freezing of local magnetic fields of different field strengths in superconductors, with the formation of the corresponding current vortex and a system of two macroscopic antiparallel vortices (of the vortex-antivortex type), has become a topic of experimental study for several research groups in recent years. The vortex pinning force corresponding to the magnetic flux quantum has been successfully measured in several film superconductors,[1] the kinetics of approach, up to their annihilation, of a frozen vortex and antivortex creating multiquantum magnetic fluxes has been observed in film superconductors,[2] and, on the contrary, the stable coexistence of such macroscopic vortex structures (MVSs) has been observed in a HTSC ceramic.[3]

This paper is devoted to a study of the features of formation and the static and dynamic parameters of the local frozen field and the corresponding MVSs in ceramic $YBa_2Cu_3O_{7-x}$. We investigate the dependence of those parameters on the manner of formation of the MVS, on the difference of the penetration of the local and uniform external magnetic fields into the superconductor, and on the features of the dynamical behavior of the MVS under the influence of a dc transport current through the slab.

Study of the properties of MVSs can potentially (perhaps by going to single-quantum frozen vortices) promote the development of new methods of physical modeling of the mixed state in superconductors by constructing vortex lattices of different degrees of complexity and may also lead to the creation of new types of weak links.

## II. DESCRIPTION OF THE EXPERIMENTS

The experimental object was a square ceramic slab with dimensions of 10 mm on a side by 0.5 mm thick. The temperature and width of the superconducting transition of the ceramic were determined, from the change in inductance of a flat coil on the surface of the slab, to be 89 and 2 K, respectively, and the critical current density at a temperature of 77 K, obtained by the standard four-probe method with the passage of a dc current through the slab, was 30 A/cm². For the formation of a MVS in the slab we used the cryogenic cell shown schematically in Fig. 1, with a working temperature of 77 K (liquid nitrogen). The cell had provisions for mounting vertically one or more composite microsolenoids (MSs) of diameter 0.5 mm. Each such MS consists of two equal parts separated by a small (0.5 mm) gap and connected in series electrically. The ceramic slab is placed horizontally in the gap. For convenience in separating out the field of the MVS from the background of the field of the Meissner current of the slab we used four pairs of MSs along the axis of the slab, with a period of 2 mm. The turns of the MSs were connected to the electrical circuit in such a way that upon passage of a dc current through the series-connected MSs, they produce in the gap local magnetic fields of equal magnitude (up to 2000 A/m) and opposite direction to the adja-cent MS. We shall call this the MVS excitation field (EF). As necessary the upper part of the MSs can be removed and the lower part mounted rigidly in the cell.

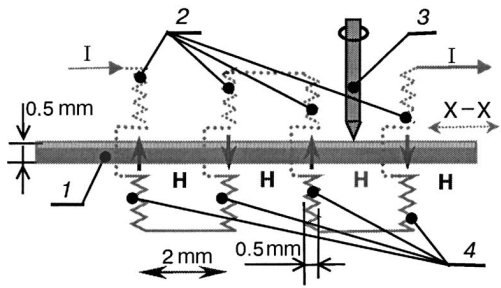

FIG. 1. Diagram of the measurements, using a magnetic detector (*3*), of the frozen magnetic field $H$ produced by a system of microsolenoids (*2*, *4*) carrying a current $I$, in a ceramic HTSC sample (*1*).

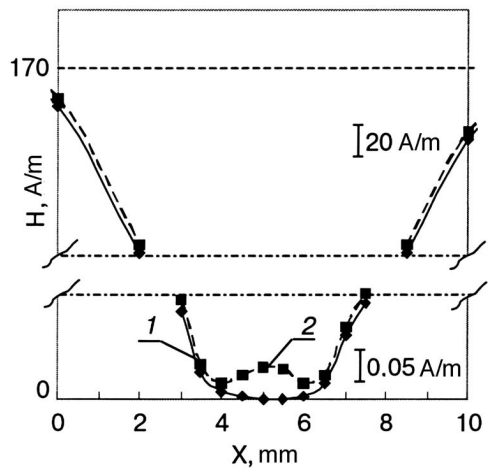

FIG. 2. Magnetic field distribution above a ceramic slab upon imposition of an external uniform field of 160 A/m (solid curve *1*) and a field of 170 A/m (dashed line *2*), at which the field begins to penetrate through the slab.

The formation of a MVS in the slab is brought about in two regimes. In the first regime the cell is cooled from room temperature to liquid nitrogen temperature (77 K) with the EF turned on (the field-cooling, FC regime) and then the EF is turned off. In the second regime the cell is cooled in the ambient magnetic field of the Earth, with the EF turned on for several seconds at 77 K to a value exceeding its critical value for the given sample and then turned off (the "zero"-field cooling, ZFC regime).

The main information about the parameters of the MVS was obtained by measuring the magnetic field above it. The value of this field and its distribution over the surface of the slab were determined by measuring its vertical component upon an axial horizontal scanning of the surface of the slab by the miniature ferroprobe detector (FD) of a magnetic microscope.[4] For this the FD was immersed in liquid nitrogen. Prior to the scanning, the four upper parts of the MSs were removed from the surface of the slab without warming of the slab. We should mention some features of the measurement of the local field of the MVS using the FD due to the design of the latter. The FD is a parametric magnetic detector in the form of an induction coil (1.5 mm in diameter and 10 mm long) with a linear ferromagnetic core on which turns of copper wire are wound uniformly along the length of the FD. The magnetic permeability of the coil varies under the influence of the field of the MVS. The sensitivity of the FD to a field which is uniform along its length is $10^{-2}$ A/m. The response of the FD to the MVS field, nonuniform along its coils, depends on the magnetic moment and shape of the fringing magnetic flux of the MVS, and also on the magnetic coupling of the MVS with the FD. The moment and fringing flux depend on the value and the MVS current distribution in the plane of the slab, while the magnetic coupling depends not only on the value and distribution of the MVS current but also on the distance of the turns of the FD coils from the MVS. At the time of the experiments the distance from the end of the FD to the surface of the slab was held fixed at around 0.1 mm.

A uniform magnetic field perpendicular to the surface of the slab was produced by a solenoid with a diameter greater than the slab dimensions.

To study the dynamic parameters of the MVS a dc transport current of regulable strength was fed to the slab. The direction of the current was chosen perpendicular to the line of MS sites and, accordingly, to the line of MVSs regions along the axis of the slab.

## III. EXPERIMENTAL RESULTS AND DISCUSSION

In the first cycle of the experiments we measured the local critical field of formation of MVSs and the uniform critical field of the sample, at which a field perpendicular to the surface of the slab begins to penetrate into the ceramic. This can also be called the Josephson critical field, since it is known that a perpendicular magnetic field first penetrates a HTSC ceramic into the spaces between grains of the ceramic, which are connected by weak links of the Josephson type. The local critical magnetic field perpendicular to the surface of the sample, $H_{c,l}$, at $T=77$ K was determined in the ZFC regime. As the value of this field we have taken the value of the EF at which, after the freezing cycle, we were able to measure a noticeable MVS field against the background noise level (around 20 A/m). It was found that $H_{c,l} \approx 1700$ A/m. Thus the fraction of the frozen field that can be registered with the FD at an EF equal to $H_{c,l}$ is only about 1% of the value of the latter.

In contrast to this, the uniform perpendicular critical field $H_{c,u}$ amounted to around 170 A/m. Here the value of the critical field was determined by the standard technique[5] with the use of the uniform-field solenoid mentioned in the previous Section. The value of $H_{c,u}$ was fixed by a detector from the first signs of the destruction of the complete diamagnetism of the ceramic slab as the uniform field is increased and the EF penetrates through it (Fig. 2). This value correlates well with the data obtained previously[5] for a similar material in a uniform field.

In the second cycle of experiments we measured the magnetic field above a MVS formed in the FC regime. The dependence of the maximum value of the field ($H_f$) of the MVS above its center on the value of the EF ($H_e$) is shown in Fig. 3. In contrast to the previous type of vortex the values of the frozen field are close to the values of the EF at all values of the latter, both below and above $H_{c,l}$. Here the dependence $H_f(H_e)$ is close to linear.

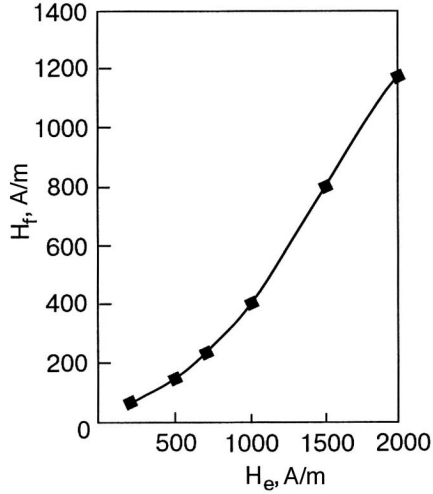

FIG. 3. Plot of the value of the magnetic field $H_f$ above the center of a frozen vortex formed in the FC regime versus the excitation field $H_e$ produced by the microsolenoid.

Finally, in the third cycle of experiments we attempted to determine the dynamic parameters of the MVS formed in the FC regime. For this purpose, after freezing of the excitation field and registering the field distribution above the ceramic slab, a dc transport current $I_t$, somewhat greater than the critical value (1.5 A), was passed through it for a short time interval (around 1 s). In the scanning of the slab a shift of the maximum of the field distribution in the direction perpendicular to the current direction was registered in the MVS region. Then a transport current was briefly passed in the opposite direction and the new MVS field distribution was registered. It was found that the maximum of the field distribution was shifted in the opposite direction. The corresponding curves of the MVS field distribution are presented in Fig. 4 for $I_t = 3$ A and a field of around 1700 A/m frozen into the MVS.

The results of the first cycle of measurements can be explained as follows.

The value of the local critical magnetic field of the ceramic is substantially greater than the critical value of the uniform field, which is explained qualitatively by the difference in the demagnetizing factors $D$ of superconducting objects of different shape. It is known[6] that the critical field $H_{c,s}$ of a sample with a shape other than a long cylinder is given in terms of the critical field $H_{c,c}$ of a long cylinder in a longitudinal field by the relation

$$H_{c,s} = H_{c,c}/(1-D). \tag{1}$$

In turn, the coefficient $D$ is determined by the ratio of the length of the object along the field to its transverse dimension. For a local field this ratio is equal to unity, while for a uniform field it is ten times less. The corresponding difference in the critical fields is observed in our experiments.

The large difference between the absolute values of the EF and the field of the MVS formed in the ZFC regime can be explained by a feature of the penetration and freezing-in of the EF in a superconducting ceramic. On the part of the slab beneath the MS the excitation field begins to penetrate into the ceramic in the those spaces between grains of the superconductor which are connected by weak links having the lowest critical currents between grains. A weak magnetic field does not penetrate into other loops having higher critical currents of the weak links nor into the grains themselves. After the EF is turned off, loops with the lower critical currents tend to maintain the magnetic flux penetrating them, and current subvortices of one direction, not exceeding the critical value, arise in them. Thus at an EF equal to $H_{c,l}$ the frozen flux in the part of the ceramic under the MS amounts to only a small fraction of the maximum possible flux produced by the EF through the whole area under the MS. As a consequence of this phenomenon, the response of the FD corresponds to a measured field significantly lower than the value of the EF. Thus the MVS, which consists of a set of microvortices, can be called a multivortex (Fig. 5a).

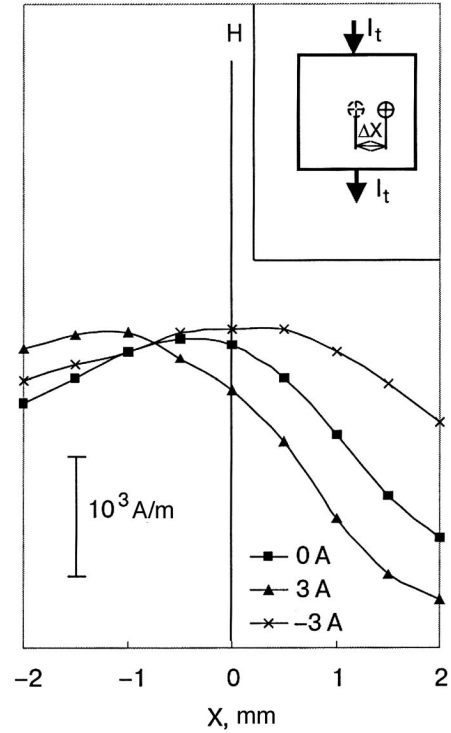

FIG. 4. Magnetic field distribution above a vortex formed in the FC regime, before (■) and after passage of a transport current $I_t$ through the samples for different values of $I_t$: 3 A (▲) and −3 A (×). The inset shows a diagram of the displacement of the vortex by a distance $\Delta X$ under the influence of the current.

It follows from the results of the second cycle of measurements that in the given case the MVS has good magnetic coupling with the FD (this follows from the approximate equality of the EF and the MVS field registered) and that nonlinear structures and processes are not involved in its formation (this follows from the linearity of $H_f(H_e)$). This indicates that in the FC regime a different type of MSV is formed than in the ZFC regime. Indeed, at a temperature near the superconducting transition $T_c$, any, even arbitrarily weak, MS field exceeds the maximum possible critical value of the superconductor at that temperature, which corresponds to a normal state of the superconductor material inside the circular area of the slab under the MS and a superconducting state in the region surrounding it. Upon further cooling of the slab a transition of the central region under the MS to the superconducting state begins, so that here the edge (on the end) field of the microsolenoid has a minimum. Here the magnetic field is expelled from the central nucleus of the superconducting phase (micron-sized ceramic grains and microloops with weak Josephson links) by virtue of the relation $\Phi_{mc}=\mu_0 H_e s < \Phi_0$, where $\Phi_{mc}$ is the magnetic flux through the grain or through the microloop, $s$ is the transverse cross-sectional area of the grain or of the microloop of comparable size, and $\Phi_0$ is the magnetic flux quantum. The diameter of the superconducting region (Fig. 5b) from which the field is expelled expands further by the same principle, since the currents that arise along the circumference of this region and along the outer circumference of the structure corresponding to the diameter of the MS, which are directed oppositely (ensuring conservation of the magnetic flux of the excitation field in the doubly connected structure that arises), strongly decrease the induction of the field in the central superconducting region. Thus a kind of positive feedback arises between an increase of the diameter of the superconducting region and a weakening of the magnetic induction of the field in the central superconducting region of the structure. Simultaneously the concentration of the excitation field in the annular region between the indicated circumferences increases. The diameter of the superconducting region stops growing when the value of these currents approaches the critical value. Then there appears inside the ring a medium with superconducting grains and nonsuperconducting weak links (in the case of their superconductivity there could be leakage and diminishing of the loop currents). The concentrated excitation field is contained in the space between grains. After it is turned off, the loop currents $i$ flowing along the indicated circumferences are frozen by virtue of magnetic flux conservation in the doubly connected superconductor (in the annular region). According to the given model, upon an increase of the EF the frozen flux and the output signal of the FD decrease proportionally, which agrees with experiment. Furthermore, the aforementioned current loops generating the frozen field have the maximum possible magnetic coupling with the FD. This is a necessary condition for proximity of the values of the EF and the frozen field registered by the FD, which also agrees with experiment.

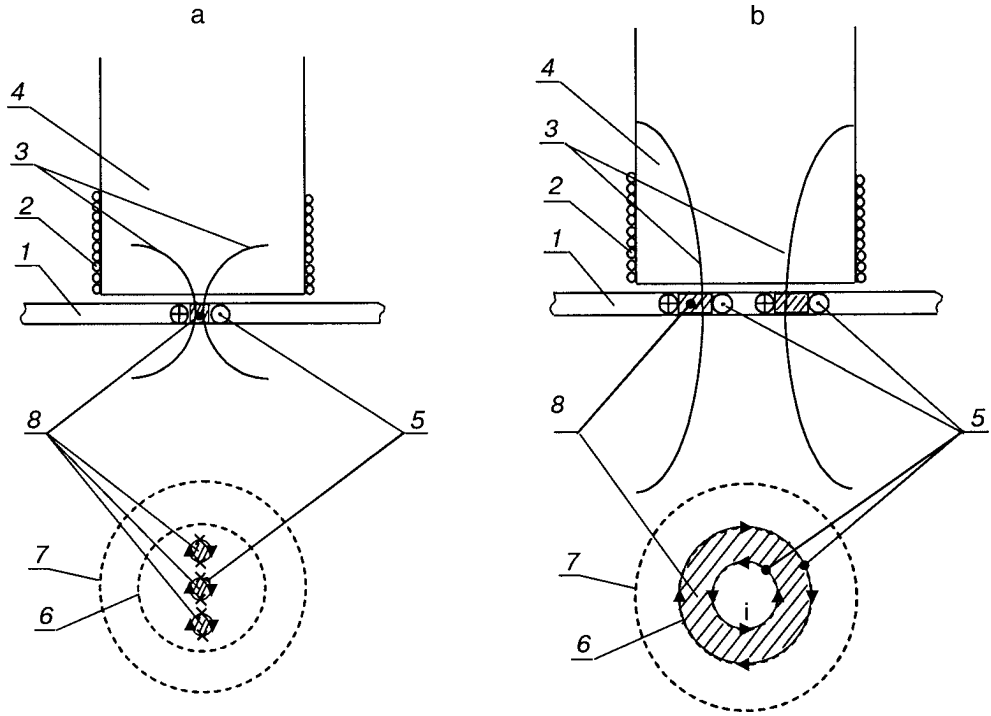

FIG. 5. Diagrams (in two projections) of the currents and fringing fields of macroscopic vortex structures: a—the structure formed in the ZFC regime (cooling in the ambient magnetic field of the Earth) and containing current subvortices of small size (only three of these are shown); b—the structure formed in the FC regime (cooling in the excitation field) and containing two coaxial current vortices (a double vortex) with diameters comparable to the diameter of the microsolenoid producing the excitation field: *1*—sample, *2*—winding of the ferroprobe detector, *3*—field lines of the fringing field of the structure, *4*—core of the detector, *5*—contours of the vortex currents, *6*—circumference of the microsolenoid, coinciding in Fig. 5b with the outer contour of the vortex current, *7*—contour of the detector, *8*—region of frozen flux.

This type of MVS with two coaxial current contours can be called a double vortex. It should be noted that, in contrast to a multivortex, a double vortex can be formed by an extremely low excitation field. The lower threshold for its formation may be the field $H_{\min}=\Phi_0/\mu_0 S_{ms}$, where $S_{ms}$ is the area of the sample under the microsolenoid producing the excitation field.

Direct observation of the magnetic microstructure of a double vortex will require a magnetic detector with higher spatial resolution than that of the FD used here.

The mechanism of formation of a double vortex suggests that such a structure can also arise upon the freezing in of a local field (in the FC regime) in other types of superconductors (nonceramic).

From the third cycle of experiments and the shift of the maximum of the distribution of the frozen field under the influence of a transport current, it may be assumed that the MVS is shifted under the influence of the Lorentz force $F_L$ caused by the interaction of the magnetic field of the MVS and the transport current. It should be noted here that at currents lower than critical, no changes whatsoever in the position of the maximum of the field distribution of the MVS were observed. Knowing the value of the frozen field of the MVS, the transport current, the time of action of the current, and the spatial shift of the maximum of the field distribution, one can estimate the pinning force and viscosity of motion of the given type of MVS. The pinning force $F_p$ was calculated from the formula[7] $F_p = F_L = j_c \Phi t$, where $j_c$ is the critical current density of the ceramic at $T=77$ K, $\Phi$ is the magnetic flux of the MVS, and $t$ is the thickness of the slab. The viscosity was estimated from the formula[8] $\eta = F_L / v_L$, on the assumption that a regime similar to the flux-flow regime exists at a current equal to twice the critical value. Here $v_L$ is the mean velocity of the shift of the maximum of the MVS field distribution at a current higher than critical. In our case $F_p = 6 \times 10^{-8}$ N, $\eta = 6 \times 10^{-5}$ kg/s.

Inasmuch as the motion of a local macroscopic current structure with a frozen field under the influence of a transport current is observed for the first time, we are planning a more detailed investigation of the motion of both types of structures under the influence of different transport currents. In view of the possible difference of the properties and structure of MVSs in the form of double and multivortices, one expects to see a difference in their dynamic parameters at different sizes and for different frozen fields.

## IV. CONCLUSION

The formation of a local frozen field (up to 2000 A/m) in a HTSC ceramic and its investigation by methods of magnetic microscopy have allowed us to establish that:

—the local critical field for penetration into a flat sample, which is the value necessary for the formation in it of a macroscopic local frozen field and the corresponding macroscopic current vortex at temperatures below the critical, can exceed many-fold the uniform perpendicular critical field of the sample; this difference can be explained by the circumstance that the demagnetizing factor of the ceramic slab studied here is much greater than the demagnetizing factor for the local region of the ceramic on which the local magnetic field acts;

—depending on the method of formation of the macroscopic vortex structures maintaining the local frozen field, at different values of their excitation field one can obtain two completely different types of structures in their magnetic parameters: one type has a relatively low value of the frozen field, apparently containing smaller subvortices within the indicated macroscopic region, while the second type, having a substantially higher frozen field, we suppose to be due to the formation of two macroscopic coaxial current vortices with oppositely directed currents of equal magnitude (a double vortex); these currents create a field in the annular region between them which is greater than the critical field of the weak links, and the outer diameter of the double vortex coincides with the diameter of the MS producing the excitation field;

—the two types of vortex structures mentioned also differ in that the first can be formed only at values of the excitation field greater than the local critical value, while the second can be formed at possibly any arbitrarily small excitation field, including a field lower than the local critical field;

—when a dc transport current (greater than the critical) is passed through a sample with a vortex structure of the second type, that vortex structure moves under the influence of the Lorentz force. Using the observed parameters of the motion, current, and frozen field of the current structure, one can directly calculate the pinning force and estimate the viscosity of motion of such a macroscopic current structure (double vortex) in the superconductor.

The authors thank Acad. I. M. Dmitrenko and Dr. A. N. Omel'yanchuk for support of this research and helpful discussions of the results.